\documentclass[twocolumn,twoside,slac_two]{revtex4}
\usepackage{graphicx}
\usepackage{fancyhdr}
\begin{document}

\title{The BES-III experiment at the high luminosity Tau-Charm factory}

%

\author{Yi-Fang Wang}
\affiliation{Institute of High Energy Physics, Beijing, 100049, P.R. China}

\begin{abstract}

Interesting results from BES-II and other experiments raised
actually many new questions which shall be answered by its upgrade
program, BEPCII and BES-III. The design and current status of
BEPCII and BES-III are reported.
\end{abstract}

\maketitle

\thispagestyle{fancy}


\section{Physics motivation}

In early 80's, Chinese government decided to build an $e^{+}e^{-}$
collider running at the tau-charm energy region, called BEPC,
which is completed in 1989. The only detector at the machine is
called Beijing spectrometer(BES). In mid 90's, there has been a
minor upgrade of the detector, which is then called BES-II. Since
then, hundreds of papers have been published on the international
journals, some with significant impacts to the community. The
upgrade of BEPC was decided at the beginning of this century,
called BEPCII, which has a designed luminosity of
$10^{33}~cm^{-2}s^{-1}$, an increase of a factor of 100. The
corresponding detector, called BES-III, adopted latest detector
technology to minimize systematic errors in order to match the
unprecedented statistics.

The physics program of the BES-III experiment includes light
hadron spectroscopy, charmonium, electroweak physics from charmed
mesons, QCD and hadron physics, tau physics and search for new
physics.  Due to its huge luminosity and small energy spread, the
expected event rate per year is historical, as listed in
table~\ref{tab:lum}.

\begin{table}[htbp]
\caption{$\tau$-Charm productions at BEPCII in one year's
running($10^7$s).} {\begin{tabular}{@{}llll}
 \hline
               & CoM energy & Luminosity     & \#Events  \\
Data Sample    & (MeV)& ($10^{33}$cm$^{-2}$s$^{-1}$)  & per year  \\
\hline
$J/\psi$ &  3097  & 0.6     & $10\times 10^9$\\
$\tau^+\tau^-$   & 3670 & 1.0  & $12\times 10^6$ \\
$\psi(2S)$ & 3686  & 1.0 & $3.0\times 10^9$ \\
$D^0\overline{D}^0$ & 3770 &1.0 & $18\times 10^6$ \\
$D^+D^-$ & 3770  &1.0 & $14\times 10^6$ \\
$D^+_S D^-_S$ & 4030  &0.6 & $1.0\times 10^6$ \\
$D^+_S D^-_S$ & 4170  &0.6 & $2.0\times 10^6$ \\
\hline
\end{tabular} \label{tab:lum}}
\end{table}

\begin{figure}
\centerline{
\includegraphics*[angle=90,width=90mm]{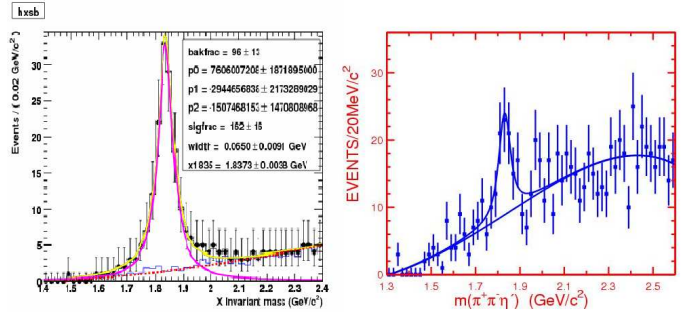} }
 \caption{A comparison of the X(1835) signal at BES-II and the
corresponding expectation at BES-III. 
(a) 2.5 days of BES-III, (b) two years of BES-II }
 \label{fig:x1835}
\end{figure}

It is well known that J/$\psi$ and $\psi$' decays is an ideal
laboratory for light hadron studies since it has a huge production
cross section with a very clean and gluon reach environment. We
plan to study the meson and baryon spectroscopy, search for
glueballs and other exotics such as hybrids and multi-quark
states. Recently, BES-II found several new structures and
threshold enhancements in various decays channels~\cite{besii},
which leads to a number of speculations. It is clear that more
data is needed to understand these results, study their decay
properties and establish a theoretical framework to accommodate
them. Fig.~\ref{fig:x1835} shows a comparison of the X(1835) signal
at BES-II and the corresponding expectation at BES-III. 
The improvement comes mainly 
from the increase of luminosity by a factor of 100, and the
decrease of the energy resolution of the electromagnetic
calorimeter by a factor of 10.  Similar results are expected for
other new structures and threshold enhancements.

Recently, a lot of new $XYZ$ resonance have been found above or
below the open charm threshold from B decays. BES-III should be
able to study the direct production of 1$^-$ states and some of
the low mass states, such as Y(2175). We will systematically study
charmonium transitions and their decays, search for rare decays
and new phenomena, and calibrate lattice QCD calculations.

For Charm physics, the precision of CKM matrix elements can be
significantly improved by measuring the leptonic and semi-leptonic
decays of charmed mesons, and test the unitarity of CKM matrix.
The D$\bar{D}$-mixing can be measured at the level of 10$^{-4}$
and the CP violation will be searched for at the level of 10$^{-3}$.
Rare and forbidden decays
can be searched for at a typical level of 10$^{-8}$.

The tau mass measurement can be improved by a factor of two over
the BES-II results with a new beam energy calibration based on 
the Compton scattering technique.

A summary of BES-III physics program, called yellow book, is under
preparation. It will be published at the beginning of next year.

\section{Status of the collider}

The new BEPCII has two storage rings with a circumstance
of 224~m, one for electron and one for positron, each with 93
bunches spaced by 8~ns~\cite{bepcii}.  The total current of the
beam is designed to be 0.93~A, and the crossing angle of two beams
22~mrad.  The peak luminosity is expected to be $10^{33}
\mbox{cm}^{-2}\mbox{s}^{-1}$ at the beam energy of 1.89 GeV,  the
bunch length is estimated to be 1.5~cm and the energy spread
$5.16\times 10^{-4}$.

At this moment, all the LINAC equipments have been installed and
successfully tested. Parameters such as the beam current, emittance
and energy spread etc. for both electron and positron beams, have
been measured. All the design specifications have been satisfied.

The storage rings have been installed in two phases. In the
first phase, conventional magnets at the interaction region were
installed and tested. Both electron and positron beams
have been stored up to 500 mA with a reasonable life time.
Synchrotron radiation beams have been delivered to user for
about three months in total. Tests of $e^+e^-$
collision have been performed with an estimated peak luminosity
of about $10^{31}~cm^{-2}s^{-1}$.
In the second phase, super-conducting quadrapoles for the final beam
focusing at the interaction region have been installed
and beams have successfully stored. Collision of $e^+e^-$ beams have
been observed and synchrotron radiation run will start soon.

We plan to move the BES-III detector into the interaction region
early next year after the luminosity is more than $3\times
10^{31}~cm^{-2}s^{-1}$ and the beam background is under control.

\section{status of the BES-III construction}

\begin{figure}
\centerline{
\includegraphics*[angle=90,width=120mm]{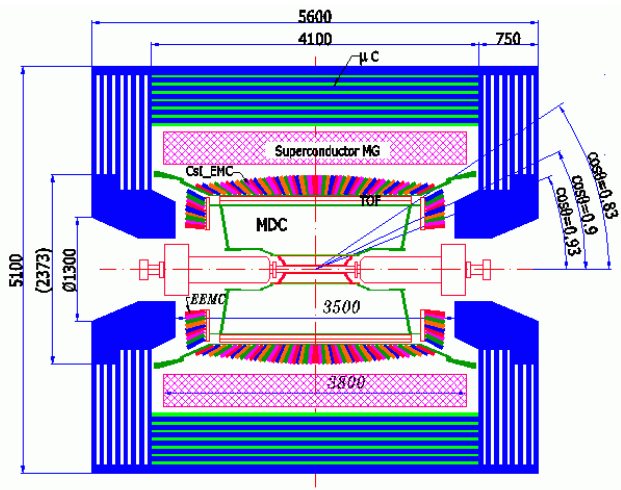}
}
 \caption{Schematics of the BES-III detector.}
 \label{fig:besiii}
\end{figure}

The BES-III detector~\cite{besiii, wang}, as shown in
Fig.~\ref{fig:besiii},  consists of a drift chamber in a small
cell structure filled with a helium-based gas, an electromagnetic
calorimeter made of CsI(Tl) crystals, Time-of-Flight(TOF) counters for
particle identification made of plastic scintillators, a muon
system  made of Resistive Plate Chambers(RPC), and a
super-conducting magnet providing a field of 1T. Current status of
the construction is summarized in the following.

The drift chamber has a cylindrical shape with two chambers
jointed at the end flange: an inner chamber without outer wall
and an outer chamber without inner wall. There are a total of 6 stepped end
flanges made of 18 mm Al plates
in order to give space for the focusing magnets. The inner radius of the
chamber is 63 mm and the outer radius is 810 mm, with a length of
2400 mm. Both the inner and outer cylinder of the chamber are made
of carbon fiber with a thickness of 1 mm and 10 mm respectively. A
total of 6300 gold-plated tungsten wires(3\% Rhenium) with a
diameter of 25 um are arranged in 43 layers, together with
a total of 22000 gold-plated Al wires for field shaping.
The small drift cell structure of the inner chamber has a dimension of $6\times6
\mbox{mm}^2$ and the outer chamber of $8\times 8 \mbox{mm}^2$,
filled with a gas mixture of 60\% helium and 40\% propane.
 The designed single wire spatial resolution and dE/dx resolution are
130~$\mu$m and 6\%, respectively.

\begin{figure}
\centerline{
\includegraphics*[angle=90,width=70mm]{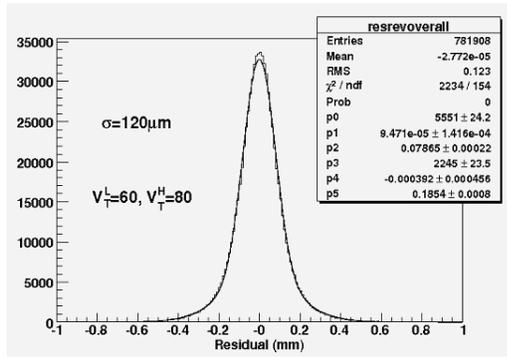}
}
 \caption{Single wire resolution of the MDC from a cosmic-ray test}
 \label{fig:mdc-cosmic}
\end{figure}

All the wiring have been completed with a very high quality, the
wire tension and the leakage current are well under control. The
assembly of the chamber has been completed together with all
preamplifiers and related electronics. The whole chamber has been
tested using cosmic-rays for three months. The obtained single
wire resolution is about 120 $\mu m$, as shown in
Fig.~\ref{fig:mdc-cosmic}, well satisfying our design goal. The
chamber has now been installed successfully into the BES-III
detector.

The CsI(Tl) crystal electromagnetic calorimeter consists of 6240
crystals, 5280 in the barrel, and 960 in two endcaps. Each crystal
is 28 cm long, with a front face of about $5.2\times 5.2
\mbox{cm}^2$, and a rear face of about $6.4\times 6.4
\mbox{cm}^2$. All crystals are tiled by 1.5$^o$ in the azimuth
angle and 1-3$^o$ in the polar angle, respectively, and point to a
position off from the interaction point by a few centimeters.
The designed energy and position resolution are 2.5\% and 6 mm at 1
GeV, respectively.

\begin{figure}
\centerline{
\includegraphics*[width=60mm]{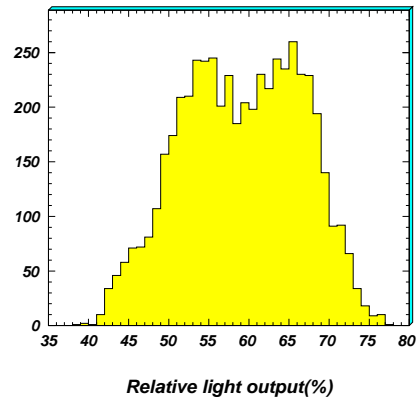}
}
\centerline{
\includegraphics*[width=60mm]{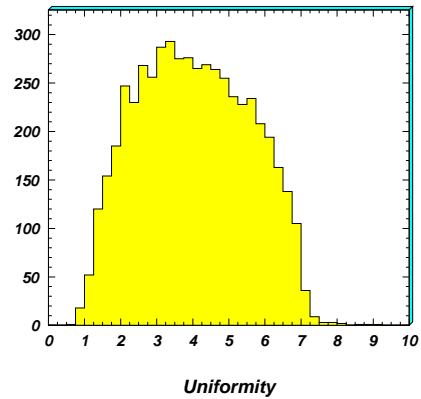}
}
\centerline{
\includegraphics*[width=60mm]{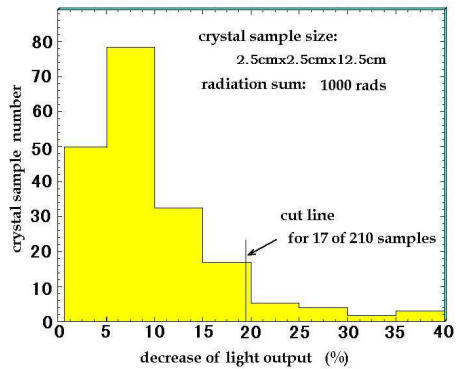}
}
 \caption{Test results of all CsI(Tl) crystals. Top: relative light yield with respect to the reference crystal; middle: Uniformity along the long axis of the crystal; bottom: radiation hardness}
 \label{fig:csi_performance}
\end{figure}

All the crystals have been produced and shipped, been tested, and assembled.
Fig.~\ref{fig:csi_performance} shows the test results of the light
yield, uniformity and radiation hardness. All the barrel crystals
have been installed into the mechanical structure, which has been
installed into the BES-III detector as well.

The readout electronics of crystals, including preamplifiers,
main amplifiers and charge measurement modules are tested at
the IHEP E3 beam line together with a crystal array and
photodiodes. Results from the beam test shows that the energy
resolution of the crystal array reached the design goal of 2.5\%
at 1 GeV and the total noise achieved the level of less than
1000 equivalent electrons, corresponding to an energy of 220 KeV.

The particle identification at BES-III is based on the momentum
and dE/dx measurements by the drift chamber, and the
TOF measurement by plastic scintillators. The
barrel scintillator bar is 2.4 m long, 5 cm thick and 6 cm wide. A
total of 176 such scintillator bars constitute two cylinders, to
have a good efficiency and time resolution. For the endcap, a
total of 48 fan-shaped scintillators form a single layer. A 2-inch
fine-mesh phototube is directly attached to each end of the scintillator 
to collect the light. The intrinsic time resolution is designed to be
90~ps including contributions from electronics and the common start/stop 
time. Such a time resolution,
together with contributions from the beam size, momentum
uncertainty, etc. can distinguish charged $\pi$ from K mesons for
a momentum up to 0.9 GeV at the 2$\sigma$ level.

Beam tests show that the intrinsic time resolution can be better
than 90~ps and 75~ps for the barrel and the endcap TOF counters,
respectively. Currently, all the PMTs and scintillators have been
delivered and tested. The average attenuation length of
all barrel scintillators is 4.8~m and the relative light yield 
exceeds our specification. All
barrel scintillator have been assembled outside of the MDC and
installed into the BES-III detector successfully.

The BES-III muon chamber is made of Resistive Plate Chambers(RPC)
interleaved in the magnet yoke. There are a total of 9 layers in
the barrel and 8 layers in the endcap, with a total area of about
$2000$ m$^2$. The readout strip is 4 cm wide, alternated between
layers in x and y directions. The RPC is made of bakelite with a
special surface treatment without linseed oil~\cite{seri}. Such a
simple technique for the RPC production shows a good quality and
stability at a low cost.
 All RPCs have been manufactured, tested, assembled and installed with
satisfaction.

The BES-III super-conducting magnet has a radius of 1.48~m and a
length of 3.52 m. It use the Al stabilized NbTi/Cu conductor with
a total of 920 turns, making a 1.0T magnetic field at a current of
3400 amp. The total cold mass is 3.6~t with a material thickness of
about 1.92~X$_0$. In collaboration with WANG NMR of California,
the magnet is designed and manufactured at IHEP.

\begin{figure}
\centerline{
\includegraphics*[angle=90,width=80mm]{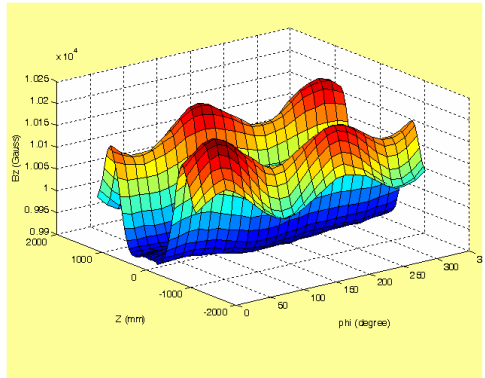}
}
 \caption{Field map inside the super-conducting magnet}
 \label{fig:field}
\end{figure}

The magnet was successfully installed into the iron yoke of the
BES-III, together with the valve box. A stable magnetic field of
1.0T at a current of 3368~A was achieved. The field mapping
together with super-conducting quadrapole magnets for final
focusing of the beam has been completed, and results are shown in
Fig.~\ref{fig:field}. The uniformity of the magnetic field is satisfactory. 

The BES-III offline software consists of a framework based on
GAUDI, a Monte Carlo simulation based on GEANT4, an event
reconstruction package, a calibration package and a database package using
MySQL. Currently all codes are working as a complete system, and
tests using cosmic-ray data and beam test data are underway. Analysis
tools such as the particle identification, secondary vertex finding,
kinematic fitting, event generator and partial wave analysis are
still under development. Data challenge of the whole system is
planed.

    In summary, the BEPCII and BES-III construction went on smoothly.
Currently all the mass production of detector components have completed,
most of the assembly and installation of the detector are finished.
We plan to take data in 2008.

\bigskip 

\end{document}